\documentclass[prd,nofootinbib,superscriptaddress,twocolumn]{revtex4}
\usepackage[a4paper,left=1.5cm,right=1.5cm,top=3cm,bottom=3cm]{geometry}

\usepackage{amssymb,amsmath,amsfonts}
\usepackage[dvipsnames]{xcolor}
\usepackage{graphicx}
\usepackage{longtable}
\usepackage{verbatim}
\usepackage{color}

\usepackage{color}
\usepackage{hyperref}
\hypersetup{colorlinks, citecolor=bluscuro, linkcolor=black, urlcolor=bluscuro}
\definecolor{rossos}{cmyk}{0,1,1,0.55}
\definecolor{bluscuro}{rgb}{0.15, 0.2, .85}
\definecolor{bluchiaro}{cmyk}{1,.3,0.,0.1}

\graphicspath{{./Figures/}}

\newcommand{\be}{\begin{equation}}
\newcommand{\ee}{\end{equation}}
\newcommand{\bea}{\begin{eqnarray}}
\newcommand{\eea}{\end{eqnarray}}
\newcommand{\beq}{\begin{equation}}
\newcommand{\eeq}{\end{equation}}
\def\beqa{\begin{eqnarray}}

\def\eeqa{\end{eqnarray}}

\def\lsim{\mathrel{\rlap{\lower4pt\hbox{\hskip0.5pt$\sim$}}
    \raise1pt\hbox{$<$}}}         
\def\gsim{\mathrel{\rlap{\lower4pt\hbox{\hskip0.5pt$\sim$}}
    \raise1pt\hbox{$>$}}}         
\def\xipbh{\xi_\text{\tiny PBH}}
\def\ppbh{P_\text{\tiny PBH}}
\def\npbh{\bar n_\text{\tiny PBH}}
\def\dpbh{\delta_\text{\tiny PBH}}
\def\mpbh{M_\text{\tiny PBH}}
\def\opbh{\Omega_\text{\tiny PBH}}
\def\rexc{x_\text{\tiny exc}}
\def\vexc{V_\text{\tiny exc}}

\def\vk{{\bf k}}

\def\vx{{\bf x}}

\def\xxi{x_\xi}
\def\keq{k_\text{eq}}
\def\kh{k_\text{H}}
\def\mh{M_\text{H}}
\def\xh{x_\text{H}}
\def\ah{a_\text{H}}
\def\aeq{a_\text{eq}}

\def\mpc{\ {\rm Mpc}}
\def\msun{\ M_\odot}
\def\mmpc{\ {\rm Mpc^{-1}}}

\usepackage[normalem]{ulem}

\newcommand{\arXiv}[2]{\href{http://arxiv.org/pdf/#1}{{\tt [#2/#1]}}}
\newcommand{\arXivold}[1]{\href{http://arxiv.org/pdf/#1}{{\tt [#1]}}}

\begin{document}

\title{Spatial Clustering of Primordial Black Holes}
\author{Vincent Desjacques }
\address{Physics department and Asher Space Science Institute, Technion, Haifa 3200003, Israel}
\email{dvince@physics.technion.ac.il}
\author{A.~Riotto}
\address{D\'epartement de Physique Th\'eorique and Centre for Astroparticle Physics (CAP), Universit\'e de Gen\`eve, 24 quai Ernest Ansermet, CH-1211 Geneva, Switzerland}
\email{Antonio.Riotto@unige.ch}

\date{\today}

\begin{abstract}
\noindent
The possibility that primordial black holes (PBHs) are the  dark matter (or a fraction thereof) has attracted much attention recently.
Their spatial clustering  is a fundamental property which determines, among others, whether current observational constraints are evaded within a  given mass range, whether merging
is significant and whether primordial black holes could generate cosmological structure.
We treat them as discrete objects and clarify the issue of their spatial clustering, with an emphasis on short-range exclusion and its impact on their large scale power spectrum.
Even if a Poissonian self-pair term is always present in the zero-lag correlation, this does not necessarily imply that primordial black holes are initially Poisson distributed.
However, while the  initial PBH clustering   depends on the detailed shape of the small-scale power spectrum, we argue that it is not relevant for a narrow spectral feature and
primordial black hole masses still allowed by observations.
\end{abstract}

\maketitle

\section{Introduction}

The physics case for the existence of Primordial Black Holes (PBHs) and for the hypothesis that all (or a fraction of)  dark matter is made of them  has regained momentum (see for instance Refs. \cite{PBH1,PBH2,PBH3,kam,kash,rep1,rep2,c0} and  \cite{revPBH} for a recent review) after the detection of  gravitational  waves generated by the  merging of two $\sim 30 M_\odot$ black holes \cite{ligo}.

In order to robustly assess whether PBHs compose a significant fraction of the dark matter, one should first investigate carefully whether current observational constraints on the PBH abundances are respected in a given mass range  \cite{Carr16}. This requires that one takes into account the possibility of PBH growth through rapid PBH merging and accretion, leading to different spatial distributions at later times \cite{ch1,ch2,g1}.
In particular, a significant clustering of PBHs could help in avoiding micro-lensing constraints as well as limits from the cosmic microwave background \cite{g2} (see, however, Ref. \cite{c1} for supernovae magnification constraints).
Furthermore, the merger rates of PBH binaries \cite{g3} along with the role played by PBHs in generating  cosmological structures depend on the clustering of PBHs \cite{carrsilk,raidal}.
Therefore, the spatial clustering of PBHs plays a crucial role but, surprisingly, little attention has been devoted to it.

The first detailed study of  the spatial clustering of PBHs at their time of formation can be traced back to Ref.~\cite{ch1} (see also Ref. \cite{jap}), in which it was argued that the PBH two-point correlation function $\xipbh(x)$ at small-scales is much larger than the   probability to form a PBH inside a horizon volume. This led the author to conclude that the mean occupation number is much larger than unity and, thus, significantly deviates from a random, Poisson distribution.
There has not been any general consensus about this conclusion  (see, for instance, the footnote of section 2.2 in Ref.~\cite{revPBH}), which has indeed been challenged very recently by Ref.~\cite{y}.

It is clear that the amount of initial clustering has an impact on the subsequent evolution and thus provides an important 
input of the problem. The  goal of this short note is not to investigate the time evolution of PBH clustering into the non-linear regime, a task which we believe can be performed satisfactorily only through suitable N-body simulations, but to clarify the issue of and offer insights on their linear clustering.
  In particular we will show that exclusion effects have a small impact on the low-$k$ PBH power spectrum, in contrast to the findings of \cite{ch1}. 
Furthermore, we will argue that, for reasonable parameters, the impact of their initial clustering on the formation of early PBH binaries can be neglected. 
This has implications as a large initial clustering makes the  constraints on PBH dark matter more severe \cite{bb}.

\section{General considerations}

PBHs may form at a given time if the energy density perturbation is sizeable enough when the corresponding wavelengths are re-entering the horizon (after inflation).
The large density contrast $\delta$ collapses to form PBHs almost immediately after horizon reentry \cite{revPBH} and the resulting  PBH mass is of the order of the mass $\mh$ contained in the corresponding horizon volume (in fact it  satisfies  a  scaling relation with initial perturbations \cite{jn}). 

We are interested in the way the PBHs are spatially distributed.
To characterize the PBH two-point correlation function $\xipbh(x)$ (or, simply, correlation function) at any comoving separation $x=|\vx|$, we can use the the peak approach to large scale structure \cite{kaiser,bbks}.
The overdensity of discrete PBH centers at position $\vx_i$ is 
\begin{eqnarray}
\dpbh({\bf x})=\frac{1}{\npbh}\sum_i \delta_D(\vx-\vx_i)-1 \;,
\end{eqnarray}
where $\delta_D(\vx)$ is the three-dimensional Dirac distribution, $\npbh$ is the average number density of PBH per comoving volume and $i$ runs over the initial positions of PBHs.
The corresponding  two-point correlation function must take the general form (see, for instance, Ref. \cite{baldauf13} in the large scale structure context)
\begin{eqnarray}
\big\langle\dpbh(\vx)\dpbh({\bf 0}) \big\rangle &=& \frac{1}{\npbh}\delta_D(\vx)-1\nonumber\\
&+&\frac{1}{\npbh^2}\Big<\sum_{i\neq j}\delta_D(\vx-\vx_i)\delta_D(\vx_j)\Big>\nonumber\\
&=&
\frac{1}{\npbh}\delta_D(\vx)+ \xipbh(x) \;.
\label{eq:PBH2pt}
\end{eqnarray}
Here, $\xipbh(x)$ is the reduced PBH correlation function. On large scales, we have
\be
\xipbh(x)=b_1^2\xi_r(x) \;,
\ee
where $\xi_r(x)$ is the radiation two-point correlation function and $b_1$ is the linear, scale-independent PBH bias.
(see \cite{revPBS} for a review on bias in large scale structure).
We will ignore the complications of higher-order biasing throughout.

As the comoving separation $x$ decreases, the ratio $\xipbh(x)/\xi_r(x)$ increases in a scale-dependent way until $x$ becomes of order the comoving size $\rexc$ of the small-scale
exclusion volume.
For most PBH scenarios, $\rexc$ is approximately  the comoving Hubble radius $\xh$ at formation time.
This spatial exclusion arises because distinct PBHs cannot form arbitrarily close to each other.
As a result, the conditional probability  $p(x|0)$ to find a PBH at a comoving distance $x$ from another PBH, which is proportional to $1+\xipbh(x)$, must vanish for $x\lesssim\rexc$. 
Therefore, the reduced correlation becomes 
\be
\xipbh(x)\approx -1 \quad \mbox{for}\quad x\lesssim \rexc \;,
\ee
that is, PBHs are anti-correlated at short distances as one should expect from the fact that there is at most one PBH per horizon volume.
This also emphasizes that, while $0\leq p(x|0)\leq 1$ surely holds, the limit $p(x\to 0|0)$ is not always unity, unlike what is assumed in Ref. \cite{y}
(see Ref. \cite{baldauf16} for counterexamples).

Finally, at zero-lag, the ``self-pairs'' contribute the well-known Poisson noise $\delta_D(\vx)/\npbh$.
This term represents the shot noise arising from the discrete nature of the PBH and, thus, is present for any distribution of point-like objects regardless of their clustering.

Therefore, while Ref.~\cite{y} properly argued that the zero-lag correlation is always of the form $\delta_D(\vx)/\npbh$ (which is equivalent to saying that PBHs are discrete
objects), this does not automatically imply that PBHs are Poisson distributed at small scales.
This will be the case only if shot noise fluctuations dominate over clustering effects.
Because power spectra and correlation functions behave differently, we shall examine this issue in both statistics.

\section{Clustering in the ``spiky'' PBH model}

Let us illustrate these general considerations with an infinitely narrow spike centered at comoving wavenumber $\kh\sim\xh^{-1}$ on top of a smooth radiation power spectrum $P_r(k)$.
We shall refer to this model as ``spiky'' PBHs.
Here, $\xh\propto\mpbh^{1/2}$ is the comoving Hubble radius when the collapsing perturbation re-enters the horizon.
This scaling holds in radiation domination, during which the mass enclosed by the horizon scales like $\mh\propto\bar\rho_r H^{-3}\propto H^{-1}$ while the comoving Hubble radius is given
by $\xh=(aH)^{-1}\propto H^{-1/2}$.
In particular, $\kh\sim 4 \times 10^6\mmpc$ for PBH of mass $\mpbh\sim\msun$. Therefore, 
\be
\label{eq:spiky}
P_r(k) = P_l(k) + \frac{2\pi^2\sigma_s^2}{k^2}\delta_D(k-\kh) \;,
\ee
where $P_l(k)$ is the power spectrum of the long-wavelength (adiabatic) fluctuations $\delta_l$ in the radiation density field, while $\sigma_s$ is the rms variance of the short-wavelength
fluctuations $\delta_s$ that collapse to form PBHs. The latter re-enter the horizon very early.
The former imprint long-wavelength fluctuations in the number density of PBH and, therefore, generates a bias analogous to the peak-background split bias of dark matter halos \cite{ch1}.

\subsection{Linear scales}

On scales $k\ll \kh$, the PBH bias is surely linear and scale-independent for the Gaussian initial conditions assumed here.
In the high peak limit (see, e.g., \cite{kaiser}), it is given by
\be
\label{eq:highpeak}
  b_1 \sim \frac{\delta_c}{\sigma_0^2(\xh)} \sim \frac{\delta_c}{\sigma_s^2}\equiv\frac{\nu_s}{\sigma_s} \;,
\ee
where $\nu_s$ is the peak significance and $\delta_c$ is the critical overdensity that leads to black hole formation.
Furthermore, $\sigma_n^2(\xh)=\sigma_{nl}^2(\xh)+\sigma_s^2 \kh^{2n}$ are the spectral moments $\langle k^{2n}\rangle$ of the radiation power spectrum smoothed on comoving scale $\xh$:
\be
\sigma_n^2(R)=\frac{1}{2\pi^2}\int_0^\infty\!\!dk\, k^2 P_r(k) W^2(kR)\;,
\ee
where $W(kR)$ is the Fourier transform of the spherically symmetric window function (assumed to be a tophat throughout).
The second approximation in Eq.(\ref{eq:highpeak}) follows from the fact that $\sigma_s^2 \kh^{2n}$ is typically much larger than the spectral moment $\sigma_{nl}^2(\xh)$ produced by the
long-wavelength piece of the radiation field solely. Therefore, the linear PBH bias (relative to the radiation distribution) will be significantly larger than unity as soon as the
variance $\sigma_s^2$ is less than $\textrm{a few}\times 0.1$.

We do not consider the case in which a PBH can be contained into a bigger PBH. In large scale structure, this is known as the cloud-in-cloud problem.
While this is not an issue for the monochromatic spike discussed here, and the narrow feature considered later, it must be taken into account when considering a broad feature.

At small scales, PBHs exclude each other because of the finite size $\xh$ of the horizons that collapse to form PBHs, as discussed above.
Upon a Fourier transform, this localized effect in configuration space affects the PBH power spectrum $\ppbh(k)$ at small $k$. 
Since the behavior of $\ppbh(k)$ in the limit $k\to 0$ is central to the argument of \cite{ch1} about the clustering of PBHs, we shall revisit it now.

\subsection{PBH shot noise at low wavenumber}

The clustering of PBHs imply that the low-$k$ white noise contribution to $\ppbh(k)$ deviates from the naive Poisson expectation $1/\npbh$.
This effect has recently been studied in the context of halo clustering \cite{hamaus,baldauf13,baldauf16,schmidt,ginzburg}.
More precisely, at small wavenumbers the PBH power spectrum is
\be
\label{lowk}
\ppbh(k) \stackrel{k\to 0}{=} \frac{1}{\npbh}+\int\!{\rm d}^3x\,\xipbh(x) \;,
\ee
which follows from the Fourier transform of Eq. (\ref{eq:PBH2pt}).
Note that this result is fully general and, thus, applies to any PBH model.

The volume integral over $\xipbh(x)$ is constrained to be larger than $-1/\npbh$, so that $\ppbh(k)$ is always positive definite. 
Furthermore, it depends sensitively on the details of the PBH clustering.
In fact, it can be positive or negative and, thus, leads to super- or sub-Poissonian noise in the low-$k$ power spectrum, respectively.
Ref.~\cite{ch1} computed the integral over $\xipbh(x)$ for ``spiky'' PBHs using the ``highly biased regions'' approximation of Ref. \cite{js} and, moreover,
truncated the integral at the lower cutoff $x=\xh$ (rather than $x=0$), finding  $\ppbh(k\to 0) > 1/\npbh$.
However, extending the integral down to $x=0$ might have reversed the sign of this integrated contribution if small-scale exclusion is significant.
This effect will not be present unless a peak constraint \cite{bbks} is enforced,
\i.e. if one requires the radiation overdensity $\delta_r$ to reach a local maximum at the PBH position.
Moreover, Ref.~\cite{ch1} extrapolated the condition $\xipbh(r)/\xipbh(0)\gg 1/\nu_s$ in regimes where it is not valid \cite{y}.

\subsection{``Halo model'' estimate of the PBH shot noise}

Although peak theory has built-in spatial exclusion, calculations are not easily tractable in this framework because the effect is highly non-perturbative
\cite[see, e.g.,][]{baldauf13,baldauf16}.
To estimate the deviation from Poisson noise in the low-$k$ PBH power spectrum, we shall therefore proceed along the lines advocated by Refs.~\cite{hamaus,ginzburg} and attempt
to establish a correspondence with the halo model approach to large scale structure.
In this framework, all dark matter is bound to halos of a wide range of mass $M$.
Consider now a narrow mass range centered at $M_i$.
At small wavenumbers, the Fourier modes $\delta_i(\vk)$ of the resulting halo fluctuation field take the general form
\be
\delta_i(\vk) = b_i \delta_m(\vk) + \epsilon_i(\vk) \;,
\ee
where $\epsilon_i(\vk)$ is the shot noise contribution.
Mass-momentum conservation implies that the mass-weighted sum of the shot noise power spectra, $\sim \sum_i M_i^2 P_{\epsilon_i}(k)$, converges towards zero in the limit $k\to 0$.
As shown in Ref. \cite{hamaus,ginzburg}, this condition can be used to obtain a consistent prediction for $P_{\epsilon_i}(k)$ from the halo model.
In plain words, one finds the following scaling
\be
\label{eq:Pshot}
P_{\epsilon_i}(k\to 0) \sim \frac{1}{\bar n_i} \left(1 - b_1 \bar n_i \vexc\right)^2 \;,
\ee
where the (comoving) exclusion volume $\vexc\equiv M_i/\bar\rho_m$ is exactly given by the Lagrangian volume of the halos.
This relation includes also the effect of the large scale clustering through the factor $b_1$.

In order to apply these considerations to PBHs, we draw the correspondence
\begin{eqnarray}
  \mbox{halos} \quad &\Leftrightarrow& \quad \mbox{PBHs} \nonumber \\
  \mbox{matter}  \quad &\Leftrightarrow& \quad \mbox{radiation} \;,
\end{eqnarray}
and take the radiation perturbation $\delta_r$ to be in synchronous gauge (comoving with the nonrelativistic matter).
This is the equivalent of the Lagrangian space used for dark matter halos \cite{hwang,wands}.
Although a fraction only  of the radiation component indeed collapses to form PBHs, the similarity with the halo model calculation is still valid. 
Therefore, on large scales, the PBH overabundance is described by 
\be
\dpbh(\vk) = b_1 \delta_r(\vk) + \epsilon(\vk)\;,
\ee
where $b_1$ is now defined relative to $\delta_r(\vk)$ instead of $\delta_m(\vk)$.
Furthermore, the shot noise contribution $P_\epsilon(k)$ is given by Eq. (\ref{eq:Pshot}), provided that $\bar n_i$ is replaced by $\npbh$ and the
characteristic volume is given by $\vexc=\mpbh/\bar\rho_r = (4\pi/3)\xh^3$.
Therefore, PBH exclusion effects on the small $k$ power spectrum will be significant when 
\be
  \label{eq:cond1}
  b_1 \xh^3 \gtrsim \bar x^3 \;.
\ee
Here, $\bar x=(3/4\pi\npbh)^{1/3}$ is the average comoving separation between PBHs. 
For the monochromatic spectrum considered here, this condition implies that the PBH linear bias be
\be
  \label{eq:cond2}
  b_1 \gtrsim \ah^{-1} \left(\frac{\Omega_r}{\opbh}\right) \;,
\ee
where $\ah$ is the scale factor at horizon entry, while $\Omega_r$ and $\opbh$ are the present-day radiation and PBH energy densities.
If PBHs make all the dark matter, Eq.~(\ref{eq:cond2}) reduces to
(assuming a number of relativistic d.o.f. $g_*(\aeq)=3.36$) 
\be
\label{eq:ah_aeq}
  b_1 \gtrsim \left(\frac{\aeq}{\ah}\right)\simeq 2.9\times 10^9 \left(\frac{\mpbh}{\msun}\right)^{-{1/2}} \;,
\ee
where $\aeq$ is the scale factor at matter-radiation equality.

We emphasize that Eq.(\ref{eq:ah_aeq}) is a criterion for the importance of a non-Poissonian correction to the low-$k$ PBH {\it power spectrum}.
The importance of the PBH {\it correlation function}, which is relevant to the formation of early PBH binaries, will be discussed below.

Clearly, the correction to Poisson shot noise in the power spectrum is completely negligible unless PBHs are simultaneously very massive
(but $\mpbh\gtrsim 10^3\msun$ is severely constrained by the data \cite{revPBH}) and strongly clustered with $b_1\gg 1$.
If all dark matter is PBHs, the white noise power induced by the PBH clustering and discreteness can thus be taken as Poissonian,
in contrast to the finding of \cite{ch1}.

We stress, however, that this is not incompatible with the fact that they can significantly cluster on sufficiently small scales. 
For the formation of PBH binaries in particular, what truly matters are the moments of the counts of neighbors (see \cite{peebles} for a review).
In particular, the mean count $\langle N\rangle$ in a cell of volume $V$ centered on a PBH is
\be
\label{eq:meancount}
\langle N\rangle = \npbh V + \npbh \int_V\!{\rm d}^3x\,\xipbh(x) \;.
\ee
$\langle N\rangle$ significantly deviates from Poisson if the contribution $\npbh\int_V\! {\rm d}^3x\,\xipbh(x)$ from PBH clustering is larger than the
discreteness noise $\npbh V$.
To quantify the importance of the former, we will compute the characteristic (comoving) clustering length
$\xxi(a)$ defined through the relation $\xipbh(\xxi(a),a)=1$.
Since the clustering is hierarchical for realistic initial power spectra, the second term in the right-hand side of Eq.(\ref{eq:meancount}) becomes
larger than the first when the cell radius is $\lesssim \xxi(a)$.

\section{Initial clustering and formation of PBH binaries}

For the formation of PBH binaries in the early Universe \cite{nakamura97}, the initial PBH clustering is the key relevant quantity \cite{ioka98,yacine2}.
The first question is whether one should compute it at horizon crossing $a=a_*$, or at the formation epoch $a=\ah$ of PBH.
Here, $a_*$ defined through $k=a_* H(a_*)$ is the scale factor at which a perturbation of wavenumber $k$ crosses the horizon.

\subsection{The initial PBH power spectrum}

We argue that the two alternatives must return the same answer because no physical process can affect the relative separation $x$ between two PBHs so
long as $x$ is larger than the horizon.
This can be captured through the requirement that $\dpbh(\vk,a)\approx b_1(a)\delta_r(\vk,a)$ remains constant for $a<a_*$,
\be
\label{eq:conservedb1}
a^2 b_1(a) = \mbox{const} \quad \mbox{for}\quad a<a_* \;.
\ee
More precisely, consider the evolution of a wavemode $\delta_r(\vk,a)$ with $k<\kh$ until it crosses the horizon at $\ah<a_*<\aeq$.
Since $k=a_* H(a_*)$ and we are in radiation domination, $a_*/\aeq=\keq/k$.
The amplitude of this mode at $a=a_*$ is given by 
\begin{align}
  \delta_r(\vk,a_*) &= \delta_r(\vk,\ah) \left(\frac{a_*}{\ah}\right)^2 \\
  &=\delta_r(\vk,\ah)\left(\frac{\aeq}{\ah}\right)^2\left(\frac{\keq}{k}\right)^2 \nonumber \;. 
\end{align}
where $\keq^2 \equiv 2 \Omega_m H_0^2/\aeq$ defines the comoving wavenumber corresponding to matter-radiation equality.
Since the radiation power spectrum at horizon crossing is
\be
P_r(k,a_*) = \left(\frac{4}{9}\right)^2 P_\zeta(k) \;,
\ee
the radiation power spectrum at black hole formation reads
\be
\label{eq:Pr_ah}
P_r(k,\ah) = \left(\frac{4}{9}\right)^2 \left(\frac{\ah}{\aeq}\right)^4\left(\frac{k}{\keq}\right)^4 P_\zeta(k) \;.
\ee
Using Eq.(\ref{eq:conservedb1}), we immediately see that
\be
b_1^2(\ah) P_r(k,\ah) = b_1^2(a_*) P_r(k,a_*) \;,
\ee
i.e. the two alternatives consistently give the same answer.
Note that these power spectra are computed before the transfer function epoch (The Meszaros effect \cite{PBH2} is thus not included).

To estimate the initial PBH correlation $\xipbh$, it is convenient to start from $\ppbh(k,\ah)=b_1^2(\ah) P_r(k,\ah)$ since PBH formation models naturally
furnish an estimate of $b_1(\ah)$.
Namely, for a peak significance $\nu_s={\cal O}(10)$ corresponding to a PBH mass fraction of the order of $10^{-9}\left(\mpbh/\msun\right)^{{1/2}}$, and a
rms variance $\sigma_s\sim 0.1$ on the PBH scale, the linear bias at PBH formation is of order $b_1(\ah)\sim {\cal O}(10^2)$ upon applying the high peak
result Eq.(\ref{eq:highpeak}).

For the ``spiky'' PBH model Eq.(\ref{eq:spiky}), with $P_l(k,\ah)$ given by Eq.(\ref{eq:Pr_ah}), the initial PBH power spectrum reads
\begin{align}
  \ppbh(& k,\ah) = b_1^2(\ah) \left[P_l(k,\ah) + \frac{2\pi^2\sigma_*^2}{k^2}\delta_D(k-\kh)\right] \nonumber \\
  &\sim 10^4 \left(\frac{b_1(\ah)}{10^2}\right)^2\bigg[10^{-39}\left(\frac{\mpbh}{\msun}\right)^{2}\left(\frac{k}{\keq}\right)^4 P_{\zeta_l}(k)
    \nonumber \\
    &\qquad + \frac{2\pi^2\sigma_*^2}{k^2}\delta_D(k-\kh)\bigg] \;,
\end{align}
where $P_{\zeta_l}(k)$ is the smooth component of the primordial curvature power spectrum.
The smallness of the numerical factor in the contribution from $P_l(k,\ah) $arises from $(\ah/\aeq)^4$. 
Furthermore, we have replaced $\sigma_s(\ah)$ in Eq.(\ref{eq:spiky}) by a free rms variance $\sigma_*$.
The reason for this choice will made be clear below.

We can now compute $\xipbh(x,\ah)$ by Fourier transforming $\ppbh(k,\ah)$.

\subsection{Clustering length}

The PBH correlation function can be expressed as
\be
\xipbh(x,\ah) = \frac{1}{2\pi^2}\int_0^\infty\!dk\,k^2 \,\ppbh(k,\ah)\,j_0(kx) 
\ee
For a scale-invariant primordial curvature power spectrum
\be
\frac{k^3}{2\pi^2} P_{\zeta_l}(k)= A_s\qquad{\rm with}\qquad A_s\sim 10^{-9} \;,
\ee
the contribution from the smooth, long-wavelength piece $P_l(k,\ah)$ of the radiation power spectrum is
(momentarily omitting the multiplicative factor of $b_1^2$ to avoid clutter)
\begin{align}
  \frac{10^{-35}}{2\pi^2}&\left(\frac{\mpbh}{\msun}\right)^{2} \int_0^\infty\!\!dk\,k^2\left(\frac{k}{\keq}\right)^4 P_{\zeta_l}(k)\, j_0(kx)
  \nonumber \\
  &=10^{-35} A_s \left(\frac{\mpbh}{\msun}\right)^{2}\int_0^\infty\!\!dk\,\frac{k^3}{\keq^4} \,j_0(kx)
  \nonumber \\
  &\simeq 10^{-35} A_s \left(\frac{\mpbh}{\msun}\right)^{2}(\keq x)^{-4}  \label{eq:xipart1} \;.
\end{align}
To obtain the last equality, we have taken advantage of the fact that the spherical bessel function satisfies $j_0(kx)\approx 1$ for $k\lesssim 1/x$,
and quickly drops to zero for $k\gtrsim 1/x$.

The calculation of the contribution to $\xipbh(x,\ah)$ arising from the spike proceeds analogously. We find
\begin{align}
  \frac{10^4}{2\pi^2}&\int_0^\infty\!dk\,k^2 \left(\frac{2\pi^2\sigma_*^2}{k^2} \delta_D(k-\kh)\right) j_0(kx) \nonumber \\
  &\simeq 10^2 \left(\frac{\sigma_*}{0.1}\right)^2 j_0(\kh x) \label{eq:xipart2}\;.
\end{align}
The PBH correlation function is the sum of Eq.(\ref{eq:xipart1}) and Eq.(\ref{eq:xipart2}).

To estimate the relative amplitude of these two terms, we note that, in the limit $x\to\infty$, the contribution arising from the smooth power spectrum
$P_l(k,\ah)$ decays as $x^{-4}$ while that arising from the spike asymptotes to a constant. Therefore, the latter clearly dominates at large scales. 
Furthermore, $x$ cannot be taken smaller than $\sim 1/\kh$. The reason is that fluctuations of wavenumber $k\gtrsim\kh$ cannot increase the {\it initial}
clustering of PBH centers because their wavelength is smaller than the size of the patches which collapse to form PBH.
In fact, those scales will contribute to decrease the amplitude of clustering through the exclusion effect discussed above. 
Therefore, the contribution from $P_l(k,\ah)$ does not exceed the upper limit (we have now restored the factor of $b_1^2(\ah)$)
\begin{align}
10^{-35} A_s &\left(\frac{b_1(\ah)}{10^2}\right)^2\left(\frac{\mpbh}{\msun}\right)^{2}\left(\frac{\kh}{\keq}\right)^4 \nonumber \\
&\simeq 10^4 A_s \left(\frac{b_1(\ah)}{10^2}\right)^2 \;,
\end{align}
where we have used the fact that $\ah\kh=\aeq\keq$. Since $A_s\sim 10^{-9}$, this upper limit is still much smaller than the contribution Eq.(\ref{eq:xipart2})
arising from the narrow feature unless the linear bias is unrealistically large, $b_1(\ah)\gg 10^2$. Therefore, the dominant contribution to the initial
PBH clustering arises from the spike itself.

We can now estimate the initial PBH clustering length $\xxi(\ah)$ upon demanding that $\xipbh(\xxi(\ah),\ah)\sim 1$, where $\xipbh$ is given by
Eq.(\ref{eq:xipart2}).
Since we want the clustering length to be significantly larger than the horizon size $\xh$, the argument of $j_0$ is $\kh\xxi\gg 1$.
Therefore, we can approximate the spherical bessel function by its envelop, $j_0(\kh\xxi)\sim (\kh\xxi)^{-1}$, to find
\be
\label{eq:xxiestimate}
\xxi(\ah) \simeq 3\times 10^{-5} \left(\frac{\sigma_*}{0.1}\right)^2\left(\frac{b_1(\ah)}{10^2}\right)^2\left(\frac{\mpbh}{\msun}\right)^{1/2}\mpc
\ee
where, again, we have restored the multiplicative bias factor for the sake of the following discussion.

\subsection{Implications}

The results are sensitive to the choice of $\sigma_*$. 
Adopting $\sigma_*\equiv \sigma_s(\ah)$ implies that we include the high density fluctuations that collapse to form PBH in the calculation of the clustering of PBH
centers. This does not seem to make sense.
In the halo model approach to large scale structure, it would be like computing the clustering of halo centers including the 1-halo term.
Therefore, these large fluctuations should be excised from the calculation, suggesting that the effective $\sigma_*$ is significantly smaller than
$\sigma_s(\ah)\sim 0.1$.

For the interesting case of PBHs with $\mpbh\sim 30 \msun$ (now severely constrained by the data if they are all the dark matter), the range of initial distances
relevant for the calculation of the present merger rate is $\gsim 4\cdot 10^{-5}$ Mpc \cite{gsm}. Our estimate Eq.(\ref{eq:xxiestimate}) indicates that $\xxi(\ah)$
can reach such values only if $\sigma_*$ is close to $\sigma_s(\ah)=0.1$.
For $\sigma_*\ll \sigma_s(\ah)$ which, as we argued above, likely reflects the contribution of the spike to the clustering of PBH centers, $\xxi$ would be much
smaller than $10^{-5}\mpc$.
Therefore, since for $\sigma_*=\sigma_s(\ah)$, $\xxi$ does not exceed significantly the bound $\gsim 4\cdot 10^{-5}$ Mpc found by \cite{gsm}, we conclude that the
initial clustering is not relevant for solar mass PBHs if they collapse out of a narrow feature in the primordial curvature power spectrum.

It is also instructive to express $\xxi$ in unit of the mean comoving PBH separation $\bar x$.
Assuming $\opbh \sim {\cal O}(0.1)$ as adopted throughout our calculation, we find
\be
\frac{\xxi(\ah)}{\bar x} \simeq 0.1\left(\frac{\sigma_*}{0.1}\right)^2\left(\frac{b_1(\ah)}{10^2}\right)^2\left(\frac{\mpbh}{\msun}\right)^{1/6}
\ee
This shows that, for a mass $\mpbh\sim 30\msun$, the condition $\xxi(\ah)\sim \bar x$ can be marginally achieved only if $\sigma_*\sim \sigma_s(\ah)$. 
Hence, our findings show that, even if the initial distribution of PBHs at formation time can potentially differ from a Poissonian, this does not happen for solar
mass PBHs and for the narrow feature considered here. For the small mass window $\mpbh\sim 10^{-12}\msun$, we find $\xxi\ll\bar x$ which shows that, here again,
the initial clustering is not significant.

We stress that the conclusions presented here do not apply to broad features. In this case, one should discard small PBHs that are quickly swallowed by bigger PBHs,
and keep only the later. This cloud-in-cloud problem could be addressed using, e.g., the excursion set approach pioneered in \cite{eps}, but this is beyond the
scope of this paper.

\section{Conclusions}

In this paper, we have discussed the basic features of PBH spatial clustering treating them as  discrete objects.
We have delineated the relation between the large-distance PBH power spectrum and short-range PBH exclusion effects.
When all dark matter is in the form of PBHs, the white noise can be taken as Poissonian, in contrast to the findings of \cite{ch1}. 
We have also emphasized that, while the zero-lag correlation includes a Poissonian self-pair contribution,
this does not mean that PBHs are  necessarily Poisson distributed at small scales. Therefore, we also disagree with some of the arguments presented in \cite{y}. 
Hierarchical clustering implies that clustering-induced fluctuations dominate on scales less than the characteristic PBH clustering length, while
Poisson fluctuations dominate on large scales. While the exact value of the  initial clustering length is  sensitive to the shape of the primordial curvature power
spectrum, our estimates suggest that, for a narrow feature, the  characteristic PBH clustering length is significantly smaller than the  mean comoving PBH
separation for resonable set of the parameters, rendering clustering not relevant.

\acknowledgments
\noindent

We thank Y. Ali-Haimoud for comments on an earlier version of this manuscript, and Chris Byrnes for a stimulating correspondence.
V.D. is grateful to Ely Kovetz and Alvise Raccanelli for helpful discussions,
and acknowledges support by the Israel Science Foundation (grant no. 1395/16). 
A.R. is supported by the Swiss National Science Foundation (SNSF), project {\sl Investigating the Nature of Dark Matter}, project number: 200020-159223.


\end{document}